\documentclass[journal=jacsat,manuscript=article, layout=twocolumn]{achemso}

\usepackage[version=3]{mhchem} 
\usepackage{upgreek}
\usepackage{xcolor}

\newcommand*\eT{$\mathrm{e_T}$\ }

\author{Angelika Demling}
\affiliation{Department of Physical Chemistry, Fritz Haber Institute of the Max Planck Society, 14195 Berlin, Germany}
\alsoaffiliation{Humboldt-Universität zu Berlin, Institut für Chemie, 12489 Berlin, Germany}
\author{Sarah B. King}
\email{sbking@uchicago.edu}
\affiliation{Department of Physical Chemistry, Fritz Haber Institute of the Max Planck Society, 14195 Berlin, Germany}
\altaffiliation{Present address: Department of Chemistry and James Franck Institute, University of Chicago, Chicago. Illinois 60637, USA.}
\author{Philip Shushkov}
\affiliation{Department of Chemistry, Tufts University, Somerville, Massachusetts 02155, USA}
\altaffiliation{Present address: Department of Chemistry, Indiana University, Bloomington, India 47405, USA}
\author{Julia Stähler}
\email{staehler@fhi-berlin.mpg.de}
\affiliation{Department of Physical Chemistry, Fritz Haber Institute of the Max Planck Society, 14195 Berlin, Germany}
\alsoaffiliation{Humboldt-Universität zu Berlin, Institut für Chemie, 12489 Berlin, Germany}

\title[An \textsf{achemso} demo]
  {\ce{O2} Reduction at a DMSO/Cu(111) Model Battery Interface}

\keywords{Lithium/Air batteries, superoxide, dimethylsulfoxide}

\begin{document}

\begin{tocentry}
\includegraphics{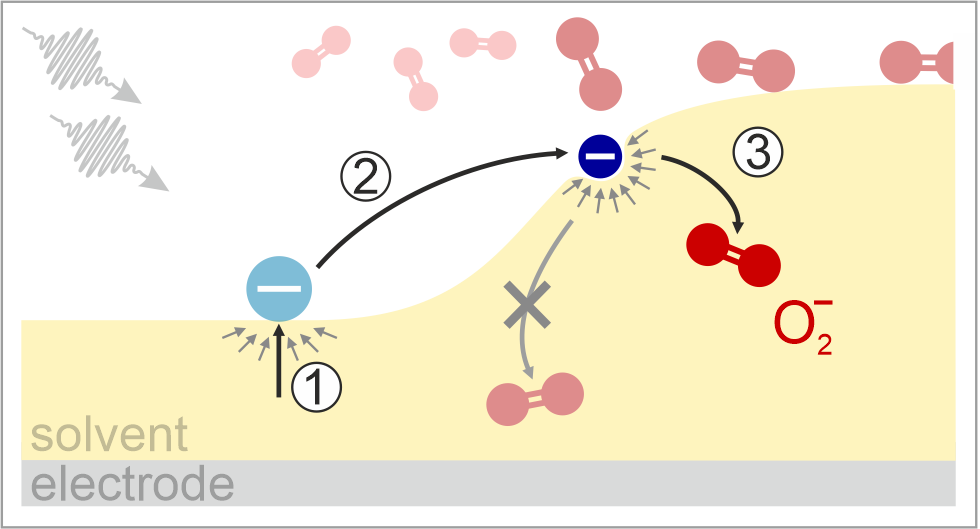}
\end{tocentry}

\begin{abstract}
In order to develop a better understanding of electrochemical $\mathrm{O_2}$ reduction in non-aqueous solvents, we apply two-photon photoelectron spectroscopy to probe the dynamics of $\mathrm{O_2}$ reduction at a DMSO/Cu(111) model battery interface. By analyzing the temporal evolution of the photoemission signal, we observe the formation of \ce{O2^-} from a trapped electron state at the DMSO/vacuum interface.
We find the vertical binding energy of \ce{O2^-} to be 3.80~$\pm$~0.05~eV, in good agreement with previous results from electrochemical measurements, but with improved accuracy, potentially serving as a basis for future calculations on the kinetics of electron transfer at electrode interfaces. 
Modelling the \ce{O2} diffusion through the DMSO layer enables us to quantify the activation energy of diffusion (31~$\pm$~6~meV), the diffusion constant (1~$\pm$~1$\cdot 10^{-8}$~cm$^2$/s), and the reaction quenching distance for electron transfer to $\mathrm{O_2}$ in DMSO (12.4~$\pm$~0.4~\AA), a critical value for evaluating possible mechanisms for electrochemical side reactions. These results ultimately will inform the development and optimization of metal-air batteries in non-aqueous solvents.
\end{abstract}
\section{Introduction}
$\mathrm{O_2}$ reduction reactions (ORR) are one half of the principle reactions of metal-air batteries, which promise extraordinarily high energy storage density from earth abundant materials. However, numerous issues hinder large scale industrial use, from competing side-reactions at the oxygen reduction electrode that form insulating, insoluble, $\mathrm{O_2}$-containing salts, to high overpotentials for oxygen reduction due to sluggish oxygen reduction kinetics.\cite{salado2022} Developing the precise solvent and electrolyte systems to promote desired reactivity and hinder side-reactions is crucial for developing metal-air battery technology, but often relies upon trial and error methods because we have a limited understanding of the fundamental distances and reaction energies relevant for oxygen reactivity at electrode interfaces.

Dimethyl sulfoxide (DMSO) is a non-aqueous solvent that has attracted attention as a solvent in lithium-,\cite{Khan2014, Liu2020, Peng2012, Johnson2014} zinc-,\cite{Hosseini2019} and sodium-air batteries,\cite{Dilimon2017} because of its role in modifying the energies of critical intermediates of the ORR. It is proposed that DMSO stabilizes \ce{O2^-} at a distance far enough from the electrode to prevent the formation of \ce{O2^{2-}} and the insulating, insoluble, $\mathrm{O_2}$-containing electrode passivation side-products.\cite{laoire2010, Johnson2014, kwabi2016} Molecular dynamics (MD) simulations predict that, immediately after the ORR, $\mathrm{O_2^-}$ is pushed approximately 10~\AA\ away from the electrode at negative cathode potentials making a second electron transfer unlikely.\cite{Sergeev2017} Determining the thickness of DMSO that prevents electron transfer to \ce{O2}, and therefore how far \ce{O2^-} needs to be away from an electrode to \emph{not} react, is critical to designing DMSO-based electrolytes for metal-air batteries, but is currently unknown and a focus of this work.

\begin{figure*}[h!]
\centering
\includegraphics[scale=1]{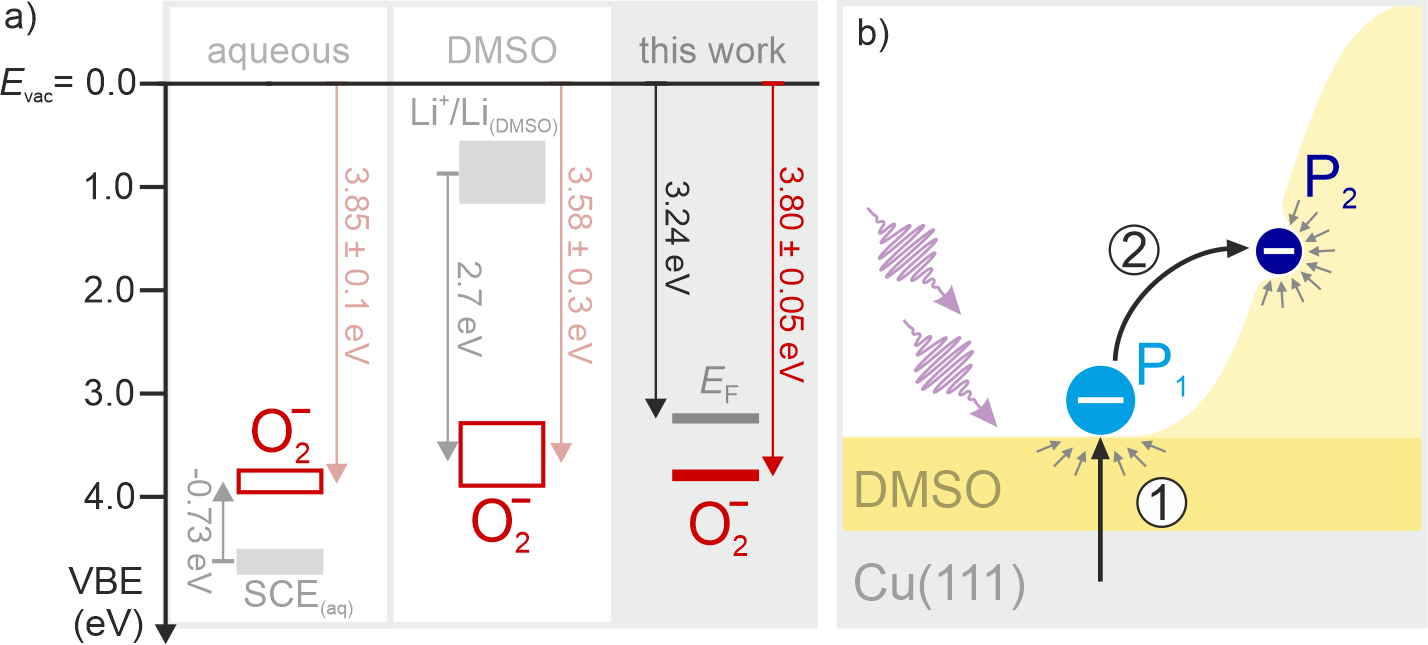}
\caption{(a) Energy level alignment of $\text{O}_2^-$ relative to $E_{\mathrm{vac}}$ based on electrochemical\cite{Trasatti1986,SAWYER196690,Peng2015,Fawcett2008} (red-bounded) and photoemission experiments (red). (b) Illustration of photoinduced polaron formation (process 1) at the surface of the two wetting monolayers of DMSO and subsequent electron trapping at the multilayer/vacuum interface (process 2). $P_1$ and $P_2$ denote the respective spectral signatures.}
\label{fig:ElChem2PPEsketch}
\end{figure*}

The Marcus theory of electron transfer has been used to predict the kinetics of electron transfer at electrode interfaces.\cite{Marcus1965, Henstridge2012, Albery1980} However, successful application of the theory requires accurate information about the energies of donor and acceptor species both before and after electron transfer, such as vertical binding energies and adiabatic electron affinities, to calculate the reorganization energy and energy of reaction that determines the interfacial reaction rate. The relevant energies for \ce{O2} and \ce{O2^-} in DMSO have been estimated from the \ce{O2}/\ce{O2^-} redox potential in DMSO measured with cyclic voltammetry (CV). The \ce{O2}/\ce{O2^-} redox potential ranges from -0.73 V in a  0.1 M \ce{(Et)4NClO4}-DMSO solution with respect to a standard calomel electrode (SCE)\cite{SAWYER196690} to 2.7 V in a 0.1 M \ce{TBAClO4}-DMSO solution versus \ce{Li}/\ce{Li^+} using \textit{in situ} surface enhanced Raman spectroscopy (SERS) measurements.\cite{Peng2015}. These redox potentials can be compared to the vacuum level of the working electrode through the standard hydrogen electrode and its vacuum level (details in the Supporting Information),\cite{Trasatti1986, Bard2001, Fawcett2008} but result in a fairly wide range of energies as shown in Figure \ref{fig:ElChem2PPEsketch}(a) (first two columns). These factors make the accurate determination of the vertical binding energy (VBE) of \ce{O2^-} in DMSO challenging and prevent accurate modeling of electron transfer reactions at electrode surfaces.

Owing to their importance in a variety of research fields, electron solvation processes have been subject of numerous studies in the ultrafast time domain applying optical and terahertz spectroscopy as well as photoemission. \cite{Alfano1993, Kimura1994, Shi1996, Knoesel2004, Bruggeman2021, Ban2021}. Two-photon photoemission (2PPE) spectroscopy of solvent molecular layers has been used to probe the energies, solvation dynamics, and reactivity of electronic states in molecular layers ranging from \ce{H2O}, \ce{NH3}, and organic solvents like \ce{DMSO}, to liquid crystals.\cite{Bertin2009, King2017, stahler2008, kwon2011, li2006,  strader2008, King2019, miller2002, muller2013} By using ultrafast laser pulses to create excited electrons above the Fermi level of a metal substrate, non-equilibrium electrons can be injected into adsorbed molecular layers and probed using a second laser pulse delayed by femtoseconds to microseconds. Using this strategy, 2PPE has been able to observe the 10s-100s of femtoseconds formation of small polarons in DMSO. In a previous publication, we discuss the electron transfer from these electronic states located on  two wetting monolayers of DMSO to much longer-living electronic states located on thicker DMSO islands with lifetimes on the order of seconds.\cite{King2019} A cartoon of the transfer process is displayed in Figure~\ref{fig:ElChem2PPEsketch}(b). Furthermore, 2PPE is capable of determining the population dynamics of such long-lived states by applying pump-wait-probe experiments (cf. Supporting Information) or measuring the repetition rate dependence of the respective feature. \cite{Bovensiepen2009, Gierster2022, Vempati2020}Even chemical reactions involving long-lived ``trapped" electrons can be detected in 2PPE experiments. For example, on water-ice surfaces, it was shown that trapped electrons can react with co-adsorbed molecules, breaking \ce{H-OH} and \ce{Cl-CCl2F} bonds and generating highly reactive hydroxide and chloride ions.\cite{King2017,Stahler2012,Bertin2009}

In this paper, we use monochromatic 2PPE of \ce{O2} adsorbed on DMSO molecular layers on Cu(111) to probe the electronic states of DMSO and \ce{O2} at a model electrode interface. We show that trapped electrons of DMSO ($P_2$ shown in Fig. \ref{fig:ElChem2PPEsketch}(b)) serve as a precursor for \ce{O2^-} formation. We measure the VBE of \ce{O2^-} to be 3.80 $\pm\ 0.05$ eV. 

By investigating the electron transfer dynamics from the trapped electron state in DMSO to $\mathrm{O_2}$, i.e. the first ORR, along with modelling $\mathrm{O_2}$ diffusion into the DMSO adlayer, we can observe the distance dependence of \ce{O2} reduction and identify the reaction quenching distance for electron transfer in DMSO. Through these experiments we have determined two components critical for accurate models of electrochemical systems with DMSO and \ce{O2}, the VBE of $\mathrm{O_2^-}$ and its formation distance, which will direct research efforts to prevent electrode passivation by unwanted oxygen reduction pathways.

\section{Methods}
The Cu(111) crystal is prepared by repeated cycles of sputtering at 0.75~kV with 1.5~$\cdot$~$10^{-6}$~mbar Ar$^+$ ions for 10 min followed by annealing at 800 K for 45 min. The surface cleanliness and order is verified by LEED, work function ($\Phi$) measurements, and the width and intensity of the surface state characteristic for Cu(111) in 2PPE spectra.\cite{Reuss1999} The $\geq$ 99.9\% anhydrous DMSO (Sigma Aldrich) is attached to the gas manifold of the ultrahigh vacuum system in an Argon atmosphere and cleaned by numerous freeze-pump-thaw cycles. Its cleanliness is confirmed by residual gas analysis. The DMSO molecules are deposited onto the copper substrate through a pinhole doser with a diameter of 5 $\upmu$m and a  backing pressure of 6~ $\cdot$~$10^{-1}$~mbar. First, molecules are deposited for 210~s with the Cu crystal temperature held at 200~K. Afterwards the sample is annealed for ten minutes at 210~K before further molecules are deposited at 150~K for 135~s, followed by a second annealing at 180~K for 10 minutes. As discussed previously\cite{King2019}, using this method a reproducible adsorption of two crystalline  DMSO monolayers partially covered with multilayer islands is achieved and the nominal layer thickness is verified with thermal desorption spectroscopy. More information about sample preparation and characterization of DMSO adlayers can be found in a previous publication and the corresponding Supporting Information. \cite{King2019} \ce{O2} molecules are adsorbed onto the DMSO/Cu(111) sample by background dosing at 46 K. At this temperature only a monolayer of \ce{O2} can be physisorbed on the surface.\cite{Dohnalek2006} All referenced temperatures are measured using a K-type thermocouple inside the copper crystal.

The laser system is a Light Conversion Pharos pump laser combined with a non-linear optical parametric amplifier (Orpheus 2H) operating at 200 kHz. This system delivers ultrashort laser pulses tunable from the visible to near UV. In the described experiments, photon energies between 2.9 eV and 3.2 eV are used, with pulse durations of approximately 100 fs.

\begin{figure*}[h!]
\centering
\includegraphics[scale=1]{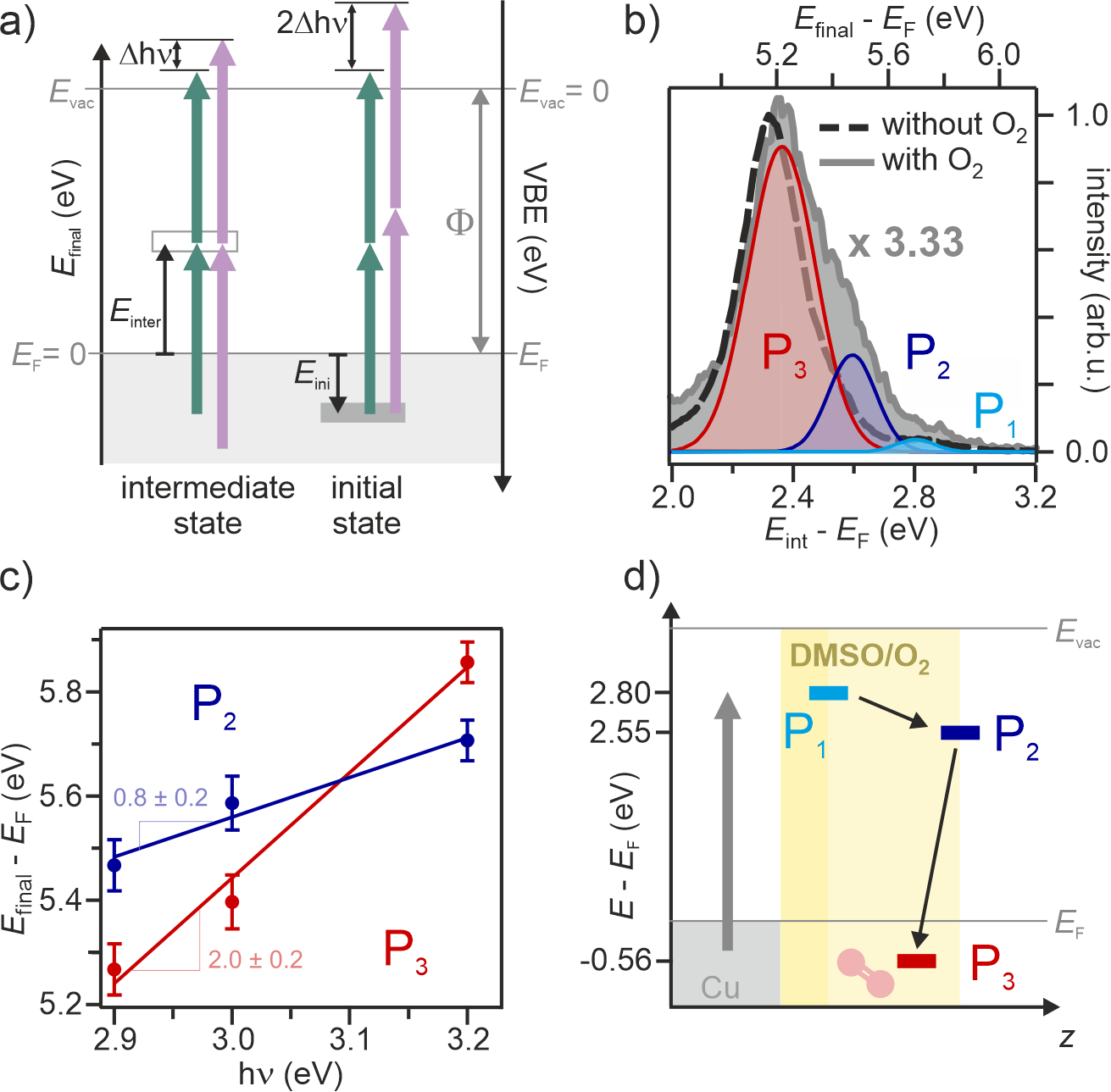}
\caption{(a) Monochromatic 2PPE schematic. The green and purple arrows show different photon energies for distinguishing intermediate and initial states in 2PPE spectra. (b) Steady-state 2PPE spectrum of 4 ML of DMSO on Cu(111) with (solid) and without (dashed) \ce{O2} adsorption. The shaded red ($P_3$), dark blue ($P_2$), and light blue ($P_1$) curves are Gaussians fitted to the 2PPE spectrum of the $\mathrm{O_2}$ dosed sample. (c) Final state energy of $P_2$ and $P_3$ versus photon energy averaged over four different fluences. The error bars refer to the standard deviations of the fits for individual measurements. (d) Energy level alignment for multilayer  DMSO  on  Cu(111) with additional \ce{O2} adsorption. After the small polaron ($P_1$, light blue) is formed, the electron can become trapped ($P_2$, dark blue) and react with \ce{O2} to form \ce{O2^-} ($P_3$, red).}
\label{fig:2PPEPhotonE}
\end{figure*}

Monochromatic 2PPE is a surface-sensitive technique that can determine the energies of both occupied and normally unoccupied electronic states with respect to the Fermi and vacuum level of the sample. The working principle is depicted in Figure \ref{fig:2PPEPhotonE}(a). In 2PPE, the absorption of two photons ionizes the sample and generates photoelectrons with finite kinetic energies with respect to the vacuum level, $E_{\mathrm{vac}}$. The kinetic energies, $E_{\mathrm{kin}}$, of the photoemitted electrons are measured with a hemispherical electron analyzer (SPECS Phoibos 100). The energy resolution of the experimental setup is better than 50 meV and determined by the spectral width of the laser pulse and the resolution of the analyzer. If the 2PPE process occurs via a real, normally unoccupied, so-called \emph{intermediate} state above the Fermi energy, $E_{\mathrm{F}}$,  variation of the photon energy by $\Delta$h$\nu$ will shift the peak by this value in the 2PPE spectra (left in Figure \ref{fig:2PPEPhotonE}(a)). If there is no real intermediate state, and the photoemission intensity results from an occupied \emph{initial} state below $E_{\mathrm{F}}$, variation of the photon energy by $\Delta$h$\nu$ will shift the 2PPE peak by $2\Delta$h$\nu$ (right in Figure \ref{fig:2PPEPhotonE}(a)). 

The energy of 2PPE features can be expressed in relation to different reference levels based on the measurement of the kinetic energy of the photoelectrons, the independent determination of $E_\text{F}$ and $\Phi$ of the sample: The final state energy, $E_{\text{final}}$ (Figure \ref{fig:2PPEPhotonE}(a), left axis), provides the excess energy of the photoelectrons with respect to $E_{\text{F}}$ 
\begin{equation}
   E_{\text{final}}=E_{\text{kin}} + \Phi
   \label{eq:fin}
\end{equation}
where $E_{\text{F}}$ is determined at a gold surface in electrical contact with the sample. $\Phi$ is given by the half-maximum of the intensity of the 2PPE spectrum's low-energy cutoff on the final state energy axis as discussed in detail previously.\cite{STAHLER2017} The intermediate and initial state energies, $E_{\text{int}}$ and $E_{\text{ini}}$, respectively, are:
\begin{equation}
   E_{\text{int}}=E_{\text{final}}-\mathrm{h}\nu
   \label{eq:inter}
\end{equation}
\begin{equation}
   E_{\text{ini}}=E_{\text{final}}-2\mathrm{h}\nu < 0
   \label{eq:initial}
\end{equation}
Finally, the VBE (Figure~\ref{fig:2PPEPhotonE}(a), right axis) with respect to $E_{\text{vac}}$ is useful when comparing energy levels to electrochemical data and is calculated by
\begin{equation}
   \text{VBE}= \Phi - E
   \label{eq:VBE}
\end{equation}
where $E$ either denotes the intermediate or initial state energy, depending on the character of the investigated state, both with respect to $E_{\text{F}} = 0$.
\section{Results}
\subsection{Energy level alignment of \mbox{$\mathrm{O_2}$-related} states}
Adsorption of \ce{O2} on top of 4 ML of DMSO on Cu(111) modifies the electronic states measured by 2PPE spectroscopy and their energy level alignment. Figure \ref{fig:2PPEPhotonE}(b) shows the monochromatic 2PPE spectra of DMSO/Cu(111) with and without \ce{O2} relative to the final (top) and intermediate state energy axis (bottom). The spectrum with \ce{O2} (grey) is notably broader to the high-energy side than the one without \ce{O2} (dashed), and can be fit with three Gaussian peaks shown in  light blue, dark blue, and red and referred to as $P_1$, $P_2$, and $P_3$, respectively. A comparison between the data and the corresponding fit is shown in the Supporting information. $P_1$ at $E_{\text{int}}$~= 2.8~eV with respect to $E_\text{F}$ is consistent with the transient small polaron in DMSO with the same energy level alignment as in the absence of \ce{O2}.\cite{King2019}

To assign $P_2$ and $P_3$ to initial or intermediate electronic states, we measured 2PPE spectra of \ce{O2}/DMSO/Cu(111) as a function of photon energy for four different photon fluences each. The peak positions of $P_2$ and $P_3$ as a function of photon energy are shown in Figure~\ref{fig:2PPEPhotonE}(c). As the photon energy is changed, $P_3$ shifts with approximately 2$\Delta$h$\nu$ while $P_2$ shifts with approximately $\Delta$h$\nu$, meaning that $P_3$ is an initial and $P_2$ is an intermediate state. Based on this, we determine the intermediate state energy of $P_2$ to be $E_{\text{int}}(P_2)$~=~2.55~$\pm$~0.03~eV and the initial state energy of $P_3$ to be $E_{\text{ini}}(P_3)= -0.56 \pm 0.03$~eV  with respect to $E_{\text{F}}$.

$E_{\text{int}}(P_2)$ is very similar to the intermediate state energy of the trapped electron of 2.34~eV observed previously\cite{King2019} on DMSO/Cu(111) \emph{without} \ce{O2}. As in this previous work, the lifetime of $P_2$ is also on the order of several seconds, as expected (cf. Introduction) and shown in the Supplementary Information. We, therefore, assign $P_2$ to a long-lived surface trapped electron ($\mathrm{e_T}$) of the \ce{O2}/DMSO surface. $E_{\text{int}}$ of $\mathrm{e_T}$ for the \ce{O2}/DMSO surface is approximately 200~meV higher in energy than for the pure DMSO surface. This suggests a destabilization of the trapped electron due to the presence of oxygen.

The VBE of $P_3$ is 3.80~$\pm$~0.05~eV, determined using equation \ref{eq:VBE} and the measured \ce{O2}/DMSO/Cu(111) work function of $\Phi= 3.24\pm0.05$ eV. We directly compare this binding energy to the approximate values of the energy of the \ce{O2}/\ce{O2^-} half reaction in DMSO obtained from cyclic voltammetry.\cite{SAWYER196690, Peng2015} Using literature values for the normal hydrogen electrode (NHE) with respect to $E_\text{vac}$ combined with the relative energies of the saturated calomel electrode (SCE) and the \ce{Li}/\ce{Li+} half reaction with respect to NHE and a correction for the use of different solvents, the CV data can be compared to $E_\text{vac}$, as shown in Figure \ref{fig:ElChem2PPEsketch}(a) and Figure~S1.\cite{Fawcett2008, Bard2001} A detailed description of how the electrode potentials are related to the vacuum potential is given in the Supporting Information. CV experiments estimate the VBE of \ce{O2^-} between 3.28 and 3.95 eV, i.e. overlapping with the VBE of $P_3$ measured in this work.\cite{SAWYER196690, Peng2015, Abraham2014} Moreover, theoretical calculations of \ce{O2^-} solvated in DMSO (Supporting Information) determine a VBE between 3.55 and 3.66 eV, in good agreement with our measured $P_3$. Based on the appearance of $P_3$ only on \ce{O2}-exposed surfaces and similarity with CV measurements and theoretical calculations, we assign $P_3$ to photoemission from \ce{O2^-}, superoxide. The differences between our measured VBE and those estimated from CV are well within the possible errors due to reference electrode to vacuum level calibration and the influence of electrolyte.\cite{SAWYER196690, Peng2015, Abraham2014}

\subsection{Mechanism of \ce{O2} reduction}
Solvent layers effectively screen adsorbed molecular species and electronic states from the metal substrate and often inhibit direct electron transfer through wave function overlap; for reactant molecules adsorbed on insulating solvent surfaces (such as DMSO), a precursor electronic state in the solvent, ideally with a long lifetime, is usually needed for electron transfer leading to anion formation from the reactant molecule.\cite{Bertin2009, King2017, King2019} In order to determine whether the short-lived small polaron or the long-lived $\mathrm{e_T}$ is a precursor state to \ce{O2^-},\cite{King2019} we took 2PPE spectra for \ce{O2}-exposed surfaces as a function of DMSO coverage. At low DMSO coverages, where $\mathrm{e_T}$ is not present, but there is a prominent signature of the small polaron,\cite{King2019} the spectroscopic signature of \ce{O2^-}, $P_3$, is not observed (not shown). Only at sufficiently high DMSO coverages above 2~ML where $\mathrm{e_T}$ is observed does the \ce{O2^-} signature appear. Therefore, we conclude that \eT is the precursor state of \ce{O2^-}, as sketched in Figure \ref{fig:2PPEPhotonE}(d), in analogy to previous work on reactivity of surface-trapped electrons at water-ice surfaces.\cite{King2017, Stahler2012} 

\begin{figure}[h!]
\centering
\includegraphics[scale=1]{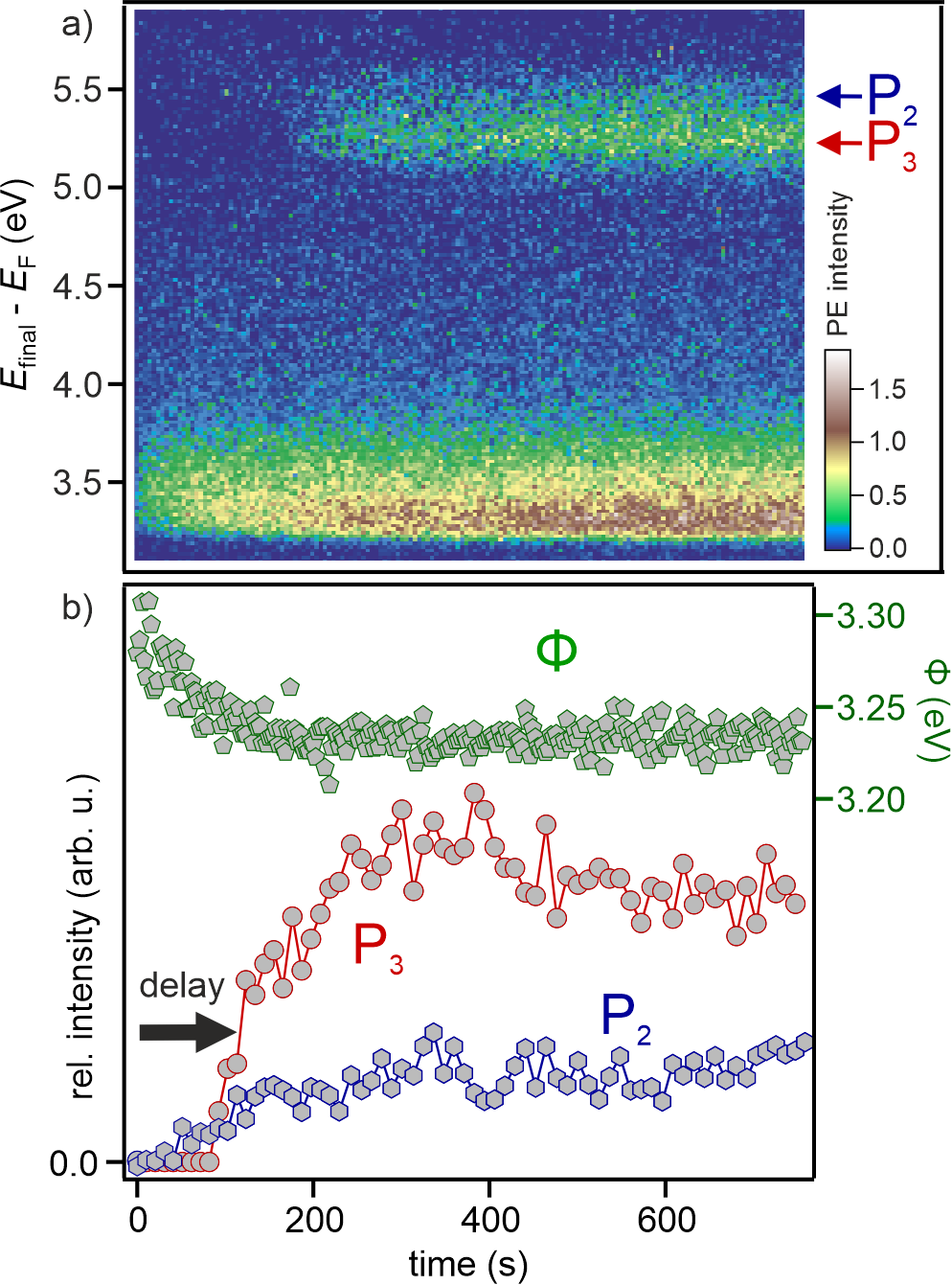}
\caption{(a) 2PPE spectra of an $\mathrm{O_2}$/DMSO/Cu(111) sample as a function of real-time illumination in false colors. The blue and red arrows indicate the positions of $P_2$ and $P_3$ signatures, respectively. (b) Evolution of $\Phi$, $P_2$ and $P_3$ intensity as a function of illumination time from the data in (a).}
\label{fig:FalseColorPlus}
\end{figure}

To determine the mechanism of \ce{O2^-} formation, we investigated the time-dependence of \ce{O2^-} formation by measuring 2PPE spectra as a function of surface illumination time. Real-time measurements, instead of ultrafast stroboscopic measurements, are required due to the lifetime of \eT that exceeds the inverse repetition rate of our laser system, 5~$\upmu$s. Figure \ref{fig:FalseColorPlus}(a) shows a representative series of 2PPE spectra of \ce{O2}/DMSO/Cu(111) as a function of illumination time in false colors measured with h$\nu$ = 2.9 eV with a photon fluence of 2.1 $\upmu$J~cm$^{-2}$. The spectral locations of \eT ($P_2$) and \ce{O2^-} ($P_3$) are shown by blue and red arrows, respectively. Both features appear after a delay of more than 100~s, in sharp contrast to the dynamics of \eT in the absence of \ce{O2}, where the trapped electron is formed on a sub-picosecond timescale.\cite{King2019} This observation is highlighted in Figure \ref{fig:FalseColorPlus}(b), which displays intensities of  \eT ($P_2$, blue) and \ce{O2^-} ($P_3$, red) extracted by Gaussian fits as in  Figure~\ref{fig:2PPEPhotonE}(b) and $\Phi$ (green) as a function of illumination time. Qualitatively, the delayed rise of $P_2$ and $P_3$ is accompanied by a decrease in the sample work function.

\begin{figure*}[h!]
\includegraphics[scale=1]{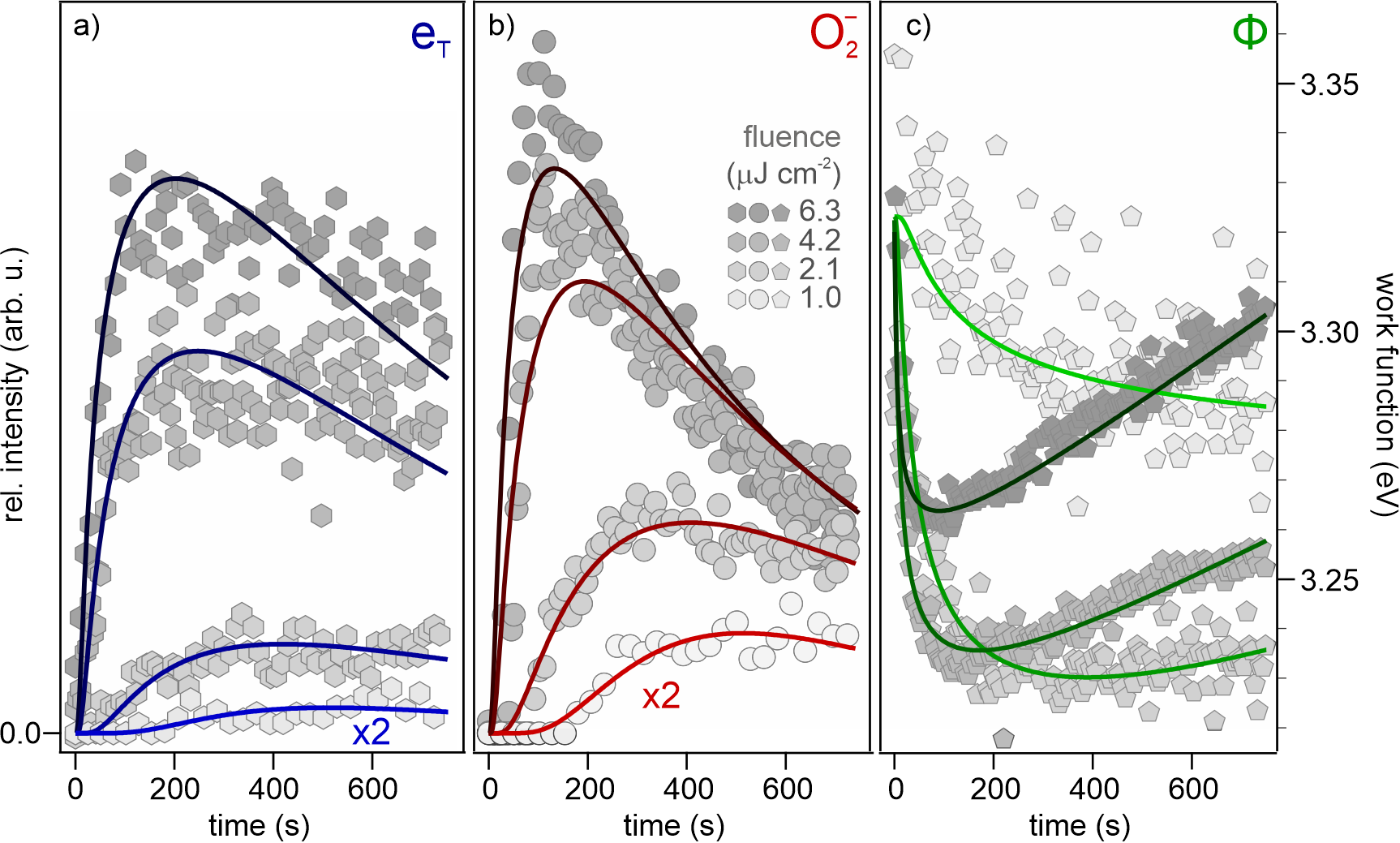}
\caption{Evolution of (a) \eT and (b) \ce{O2^-} intensities, and (c) $\Phi$  measured with h$\nu$ = 2.9 eV for four different fluences. The lines denote the fit results from a global fit described in the text.}
\label{fig:FluenceDep}
\end{figure*}

In order to explore the origin of these unusual dynamics, we collected similar datasets for different photon energies and fluences, all of which can be found in the Supporting Information. We plot representative data for the fluence-dependent dynamics of \eT, \ce{O2^-}, and $\Phi$ for h$\nu=2.9$~eV in Figure \ref{fig:FluenceDep}. As the fluence is increased, the populations of \eT and \ce{O2^-} increase, they appear at earlier illumination times, and the early-time drop in the work function occurs faster.

\subsubsection{Model of \ce{O2} diffusion in DMSO}
We propose that the increase in amplitude of \eT and \ce{O2^-}, the decrease in their appearance time, and the faster $\Phi$ decrease observed with increasing photon fluence are all due to diffusion of \ce{O2} into the DMSO layers and reactivity of \eT and \ce{O2} to form \ce{O2^-}. We will now go through this proposed mechanism in detail.

\eT in DMSO are surface-bound electronic states, likely localized at defects within the topmost monolayer.\cite{King2019} Adsorption of gases on the surface leads to a decrease and/or quenching of photoemission from these states due to dielectric squeezing or blocking of trapped electron binding sites. \cite{Bertin2009, Stahler2007, King2019} At the same time, additional adsorbates may alter the surface potential and decrease or enhance the sample work function. The delayed appearance of $\mathrm{e_T}$, accompanied by the work function decrease, suggests that something quenching \eT photoemission intensity and increasing the work function has been removed. As these observations only occur when \ce{O2} is adsorbed, it seems evident that molecular oxygen is the source of both observations. 

\begin{figure*}[h!]
\centering
\includegraphics[scale=1]{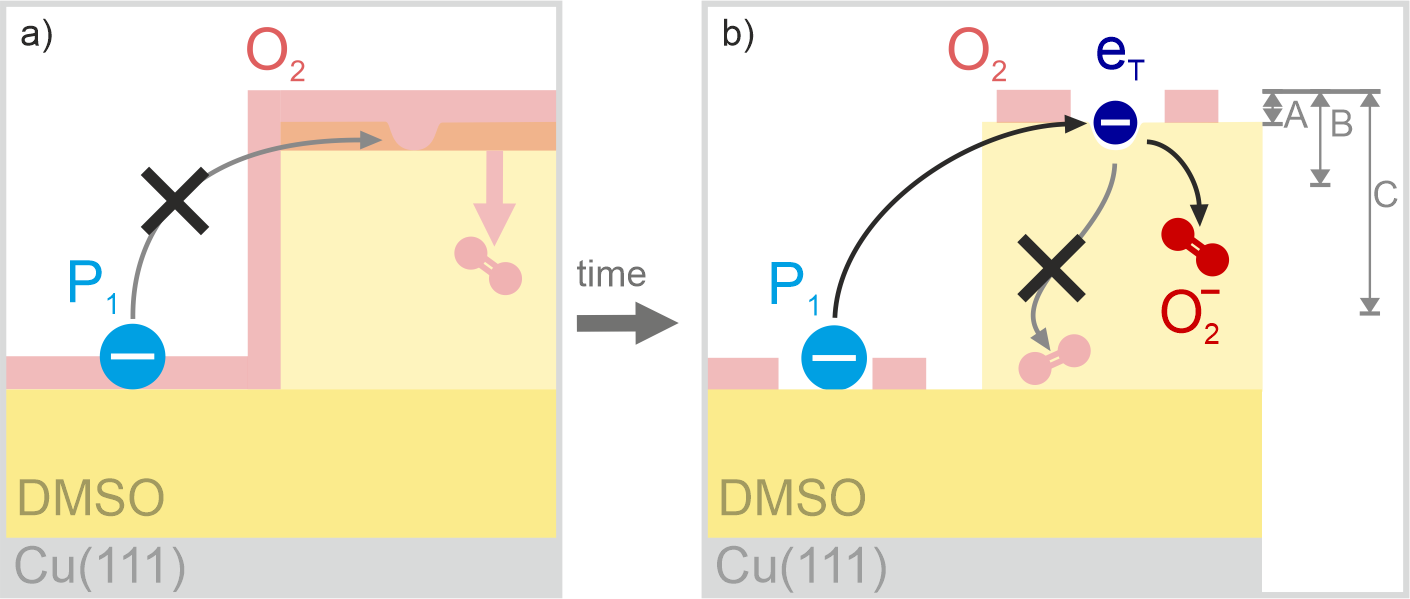}
\caption{(a) Sketch of sample before illumination, T = 46 K. $\text{O}_{2}$ induces an interfacial dipole and blocks the trapping sites. Hence, no $\mathrm{e_T}$ formation is possible. (b) Sketch of sample under illumination, T $>$ 46 K. $\text{O}_{2}$ diffuses through the adsorbed DMSO. $A$, $B$ and $C$ refer to different regions in DMSO as described in the text.}
\label{fig:SampleSketch}
\end{figure*}

Both, desorption of \ce{O2} from the surface or diffusion into the DMSO layers - driven by heat supplied by the ultrafast laser system - could be the source of surface \ce{O2} removal and lead to the delayed \eT intensity rise and work function decrease. 
However, if desorption were the cause, the spectrum after the delayed rise should coincide with a spectrum where \ce{O2} was desorbed by heating the copper substrate above the desorption temperature of \ce{O2} but below the one of DMSO. As shown in Figure~S5 in the Supporting Information, this is not the case. Hence, we conclude that diffusion of \ce{O2} into the DMSO layers is the cause for the appearance of \eT (cf. illustration in Figure~\ref{fig:SampleSketch}).
In addition, decrease in the surface concentration of \ce{O2} through diffusion would also explain the drop in $\Phi$ that occurs on the same timescales as \eT and \ce{O2^-} appearance.

We are able to capture the \ce{O2^-} photoemission intensity changes with a simple model of \ce{O2} diffusion in DMSO as follows. 
The diffusion of the \ce{O2} layer into DMSO is approximated using Fick's law of diffusion in one dimension where all diffusion occurs from the surface into DMSO. In most solids, the diffusion constant $D(T)$ follows an Arrhenius law
\begin{equation}
    D(T)=D_0\cdot\mathrm{exp}\left(-\frac{E_{\text{diff}}}{k_{\text{B}}T}\right)
    \label{Arrhenius}
\end{equation}
where the diffusion depends on an effective temperature, which in our experiments is determined by the photon fluence and is higher than the equilibrium temperature of the sample. We simplify the system into three regions shown graphically in Figure \ref{fig:SampleSketch}(b): the monolayer of adsorbed oxygen $A$ that contributes to the formation of an interfacial dipolar layer and modifies the work function, the region $B$ where near-surface oxygen can still block electron trapping sites, and the reaction region $C$ where the distance of \ce{O2} from the surface is small enough that there is sufficient wave function overlap for electron transfer from \eT to \ce{O2}. Once the \ce{O2} molecules have diffused beyond $C$, they cannot be reduced anymore, and the reaction is quenched.

The photoemission signals from \eT and \ce{O2^-}, $P_2$ and $P_3$, respectively, can both be fit using differential equations that account for the photoexcitation and photoemission processes, diffusion of oxygen and reaction of the trapped electron with \ce{O2} forming \ce{O2^-}. Both are proportional to the incident laser fluence, $F$, and the time-dependent number of accessible sites for \eT formation, $N_{acc}(t)$ 

\begin{equation}
    I_{\mathrm{e_T}} \propto F \cdot N_{acc}(t)
    \label{eT_intens}
\end{equation}

\begin{equation}
    I_{\mathrm{O_2^-}} \propto F \cdot N_{acc}(t) \cdot n_{\mathrm{C}}(t)
    \label{O2_intens}
\end{equation}

where the proportionality constants are determined by the different cross sections for photoexcitation and photoemission, and $n_{\mathrm{C}}(t)$ describes the number of \ce{O2} molecules within the region $C$. The density of accessible sites for \eT formation 

\begin{equation}
    N_{acc}(t) = N_S(t)\cdot\bigg(1-\frac{n_B(t)}{n_{O_2}}\bigg)
    \label{accessible}
\end{equation}

depends on the density of electron trapping sites $N_S(t)$, which decreases exponentially with time due to healing of surface defects as observed previously for \eT on crystalline $\mathrm{D_2O}$ on Ru(001)\cite{Gahl2004}. The ratio $n_B(t)/n_{O_2}$ describes the fraction of oxygen molecules in the region $B$ that contribute to blocking of trapping sites, i.e. $N_{acc}(t)$ is the fraction of unblocked trapping sites for \eT.

The time-dependence of $\Phi$ is fit by considering the number of oxygen molecules $n_A(t)$ that remain in the surface region $A$  

\begin{equation}
    \Delta \Phi(t) = \mu \cdot n_A(t)
    \label{WF}
\end{equation}
where each surface oxygen molecule contributes a small interfacial dipole $\mu$ that modifies the work function. All of the equations used in the fitting and their detailed explanations can be found in the Supporting Information.

Using this simple diffusion model, we can globally fit the dynamics of \eT and \ce{O2^-} across all of the fluences and photon energies. $\Phi$ is also included in the global fit at 2.9 eV, but not at 3.0 and 3.2 eV, as the early-time work function changes at those photon energies occur too quickly to be captured in our experiments, as detailed in the Supporting Information. As shown in Figure \ref{fig:FluenceDep} and Figure~S7, the global fit shows an excellent agreement with the data. From this, we extract the activation energy of diffusion,  $E_\text{diff}~=~31\pm6$~meV and the temperature-independent diffusion constant of \ce{O2} in solid DMSO, $D_0 = (1\pm1)$~$\cdot$~$10^{-20}$~cm$^2/$pulse,  which can be translated to $D_0 = (1\pm1)\cdot 10^{-8}$~cm$^2/$s assuming that each laser pulse heats the surface for approximately 1~ps until the heat is dissipated in the metal substrate through electron-electron and electron-phonon scattering \cite{Fann1992}. To our knowledge, this is the first time $D_0$ and $E_\mathrm{diff}$ of $\mathrm{O_2}$ diffusion in solid DMSO has been measured. For $\mathrm{O_2}$ diffusion in amorphous solid water, similar values were found \cite{He_2018}.

In the diffusion model, the regions $A$, $B$, and $C$ are dependent upon one another, one region must be set. $A$ is the region where the oxygen interfacial dipole modifies the sample work function, which we set to the van der Waals radius of an oxygen atom, 1.54 \AA,\cite{Mantina2009} as a reasonable minimum distance of the width of the one monolayer of oxygen molecules that forms on the DMSO surface at our deposition temperature of 46 K. Based on this, our global fit yields $B= 4.4\pm0.1$~\AA\ and $C = 14.0\pm0.4$~\AA. From these distances we can calculate the DMSO thickness needed to prevent \ce{O2} blocking of \eT trapping sites, $B-A$, is $2.8\pm0.6$~\AA, and the thickness of DMSO needed to prevent electron transfer to \ce{O2}, $x_\text{react}=C-A$, is $12.4\pm0.4$~\AA. Since the distances $B$, $C$, and $x_\text{react}$ are proportional to $A$, the extracted values from the modelling must also be considered as minimum distances.

We can compare these values with other known distances for DMSO solvation. The radial pair distribution function obtained from X-Ray diffraction and molecular dynamics calculations of solid and liquid DMSO describes how DMSO organizes around other DMSO molecules or around solutes, such as \ce{O2} and \eT studied here. The first coordination shell of liquid DMSO is found between 3-5 \AA\ from other DMSO molecules and solutes.\cite{Skaf1997, Sergeev2017} Consistent with this, our experiments found that when \ce{O2} was within $B-A$~=~$2.8\pm0.6$ \AA, from the surface, i.e. within the first solvation shell of surface-trapped electrons \eT,  it could suppress the \eT signal. Furthermore, the end of the third coordination shell of DMSO is approximately 11-12 \AA.\cite{Skaf1997, Sergeev2017} This value is in good agreement with the reaction quenching distance, $x_\text{react}$, that we determined to be the thickness of DMSO that inhibits electron transfer to oxgyen. In other words, on average three solvation shells of DMSO are required to sufficiently damp the electron wave function (i.e. screen the excess electron) to prevent electron transfer from an electronic state such as the \eT in our experiments. It seems plausible, that similar  reaction quenching distances would apply for electron transfer from an electrode surface to \ce{O2} or \ce{O2^-}.

The reaction quenching distance for electron transfer in DMSO helps determine the possible mechanisms for the formation of unwanted side-products in metal-air batteries. Simulations by Sergeev et al. \cite{Sergeev2017} found that \ce{O2^-} sits between 10-12 \AA\ away from the cathode, far enough away that our experiments suggest DMSO could significantly suppress electron transfer. In contrast, \ce{LiO2}, which has a high concentration 7~\AA\ from the cathode surface, \cite{Sergeev2017} is unlikely to be protected from further reduction by DMSO alone because, from our experiments, 7~\AA\ of DMSO are insufficient to prevent electron transfer by screening. Our experiments therefore corroborate the conclusion of Sergeev et al.\cite{Sergeev2017} that reduction of \ce{LiO2} to \ce{LiO2^-} is a more likely source of electrode passivation in Li-Air batteries than \ce{O2^{2-}} formation and suggests that the screening of DMSO plays a significant role in this mechanism.

\section{Conclusion}
In conclusion, we report the formation of $\text{O}_2^-$ near a DMSO/Cu(111) model battery interface. We were able to directly measure the VBE of $\text{O}_2^-$ solvated by DMSO is 3.80 $\pm$ 0.05 eV with great accuracy, which may serve as a basis for future calculations on the kinetics of electron transfer at electrode interfaces. We also determined the energy barrier of oxygen diffusion and diffusion constant in solid DMSO, the reaction quenching distance, $x_\text{react}$, of electron transfer in DMSO, which suppresses $\mathrm{O_2}$ reduction after approximately 12 \AA\ (approximately three solvation shells of DMSO). Optimizing the electrolyte system to allow for optimal \ce{O2^-} formation but suppress side reactions such as \ce{O2^{2-}} and \ce{LiO2^-} formation could prevent Li-air battery electrode passivation, a limitation to their repeated use. Our experiments can inform the selection of solvents for battery electrolytes in general by demonstrating the degree of screening required to prevent reduction between diffuse electronic states and small molecular species in solution, an important design criterion.

\section{Supporting information}
\begin{itemize}
    \item Conversion between electrochemical potentials and the vacuum level
    \item Shuttering experiments and decrease in trapped electron lifetime by \ce{O2}
    \item Fit of steady-state 2PPE
    spectrum from \ce{O2}/DMSO/Cu(111)
    \item Diffusion vs. desorption of $\mathrm{O_2}$
    \item Derivation of diffusion model
\end{itemize}
\section{Acknowledgment}
S.K. acknowledges funding by the Alexander-von-Humboldt-Foundation of a postdoctoral fellowship.



\bibliography{REF}

\end{document}